\documentclass[12pt]{article}
\usepackage{amsfonts}
\begin{document}
\title{\bf{A quantum long time energy red shift:\\
a contribution to varying $\alpha$ theories}}
\author{\hfill \\ K. Urbanowski\footnote {e--mail:
K.Urbanowski@proton.if.uz.zgora.pl; K.Urbanowski@if.uz.zgora.pl}
\\  \hfill  \\
University of Zielona G\'{o}ra, Institute of Physics, \\
ul. Prof. Z. Szafrana 4a, 65--516 Zielona G\'{o}ra, Poland.}
\maketitle

\begin{abstract}
By analyzing the survival probability amplitude of an unstable state
we show that the energy corrections to this state in the long ($t
\rightarrow \infty$) and relatively short (lifetime of the state)
time regions, are different. It is shown that in the considered
model the above corrections decrease to zero as $t \rightarrow
\infty$. It is hypothesized that this property could be detected by
analyzing the spectra of distant astrophysical objects. The above
property of unstable states may influence the measured values of
possible deviations of the fine structure constant $\alpha$ as well
as other astrophysical and cosmological parameters.
\end{abstract}
\section{Introduction}
Within the quantum theory the state vector at time $t$, $|\Phi
(t)\rangle$, for the physical system under consideration which
initially (at $t = t_{0} =0$) was in the state $|\Phi\rangle$   can
be found  by solving the  Sch\"{o}dinger equation
\begin{equation}
i\hbar \frac{\partial}{\partial t} |\Phi (t) \rangle = H |\Phi
(t)\rangle, \;\;\;\;\; |\Phi (0) \rangle = |\Phi\rangle, \label
{Schrod}
\end{equation}
where $|\Phi (t) \rangle, |\Phi \rangle \in {\cal H}$,  ${\cal H}$
is the Hilbert space of states of the considered system, $\| \,|\Phi
(t) \rangle \| = \| \,|\Phi \rangle \| = 1$ and $H$ denotes the
total selfadjoint Hamiltonian for the system. If one considers an
unstable state $|\Phi \rangle \equiv |\phi\rangle$ of the system
then using the solution $|\phi (t)\rangle$ of Eq. (\ref{Schrod}) for
the initial condition $|\phi (0) \rangle = |\phi\rangle$ one can
determine the decay law, ${\cal P}_{\phi}(t)$ of this state decaying
in vacuum
\begin{equation}
{\cal P}_{\phi}(t) = |a(t)|^{2}, \label{P(t)}
\end{equation}
where $a(t)$ is  probability amplitude of finding the system at the
time $t$ in the initial state $|\phi\rangle$ prepared at time $t_{0}
= 0$,
\begin{equation}
a(t) = \langle \phi|\phi (t) \rangle . \label{a(t)}
\end{equation}
We have
\begin{equation}
a(0) = 1. \label{a(0)}
\end{equation}
From basic principles of quantum theory it is known that the
amplitude $a(t)$, and thus the decay law ${\cal P}_{\phi}(t)$ of the
unstable state $|\phi\rangle$, are completely determined by the
density of the energy distribution $\omega({\cal E})$ for the system
in this state \cite{Fock},
\begin{equation}
a(t) = \int_{Spec.(H)} \omega({\cal E})\;
e^{\textstyle{-\frac{i}{\hbar}\,{\cal E}\,t}}\,d{\cal E}.
\label{a-spec}
\end{equation}
where $\omega({\cal E}) > 0$.

Note that (\ref{a-spec}) and (\ref{a(0)}) mean that there must be
\begin{equation}
a(0) = \int_{Spec. (H)} \omega ({\cal E})\,d{\cal E} = 1.
\label{a(0)-spec}
\end{equation}
From the last property  and from the Riemann--Lebesgue Lemma it
follows that the amplitude $a(t)$, being the Fourier transform of
$\omega ({\cal E})$ (see (\ref{a-spec})),  must tend to zero as $t
\rightarrow \infty$ \cite{Fock}.

In \cite{Khalfin} assuming that the spectrum of $H$ must be bounded
from below, $(Spec.(H)\; > \; -\infty)$, and using the Paley--Wiener
Theorem \cite{Paley} it was proved that in the case of unstable
states there must be
\begin{equation}
|a(t)| \; \geq \; A\,e^{\textstyle - b \,t^{q}}, \label{|a(t)|-as}
\end{equation}
for $|t| \rightarrow \infty$. Here $A > 0,\,b> 0$ and $ 0 < q < 1$.
This means that the decay law ${\cal P}_{\phi}(t)$ of unstable
states decaying in the vacuum, (\ref{P(t)}), can not be described by
an exponential function of time $t$ if time $t$ is suitably long, $t
\rightarrow \infty$, and that for these lengths of time ${\cal
P}_{\phi}(t)$ tends to zero as $t \rightarrow \infty$  more slowly
than any exponential function of $t$. The analysis of the models of
the decay processes shows that ${\cal P}_{\phi}(t) \simeq
e^{\textstyle{- \frac{\gamma_{\phi}^{0} t}{\hbar}}}$, (where
$\gamma_{\phi}^{0}$ is the decay rate of the state $|\phi \rangle$),
to an very high accuracy  for a wide time range $t$: From $t$
suitably latter than some $T_{0} \simeq t_{0}= 0$ but $T_{0} >
t_{0}$ up to $t \gg \tau_{\phi} = \frac{\gamma_{\phi}^{0}}{\hbar}$
and smaller than $t = T_{\infty}$, where $T_{\infty}$ denotes the
time $t$ for which the nonexponential deviations of $a(t)$ begin to
dominate (see eg., \cite{Khalfin}, \cite{Goldberger} --
\cite{Greenland}). From this analysis it follows that in the general
case the decay law ${\cal P}_{\phi}(t)$ takes the inverse
power--like form $t^{- \lambda}$, (where $\lambda
> 0$), for suitably large $t \geq T_{\infty}\gg \tau_{\phi}$
\cite{Khalfin}, \cite{Goldberger} -- \cite{Peres}. This effect is in
agreement with the  general result (\ref{|a(t)|-as}). Effects of
this type are sometimes called the "Khalfin effect" (see eg.
\cite{Arbo}).

Note that one can force ${\cal P}_{\phi}(t)$ to have an exponential
form for all $t$ including $t \rightarrow \infty$ by using the time
dependent decay rate \cite{Arbo,PRA}.

Unfortunately, there is no experimental evidence for the Khalfin
effect. The performed  test of the decay laws did not indicate
deviations from the exponential form of ${\cal P}_{\phi}(t)$ at the
long time region \cite{Wessner,Norman1,Greenland}, neither did they
show a time dependence of the decay rate at this time region.
Following the estimations given in \cite{Newton} one infers that
this probably occurs because it is almost impossible to build in a
laboratory a device detecting decays of the unstable states in the
vacuum which allows one to observe these decays from the instant $t
= t_{0} =0$ of the preparation of these states up to times $t \sim
10^{n}\,\tau_{\phi}$, where $n \geq 9$.

Note that in fact the amplitude $a(t)$ contains information about
the decay law ${\cal P}_{\phi}(t)$ of the state $|\phi\rangle$, that
is about the decay rate $\gamma_{\phi}^{0}$ of this state, as well
as the energy ${\cal E}_{\phi}^{0}$ of the system in this state.
This information can be extracted from $a(t)$. Indeed if
$|\phi\rangle$ is an unstable (a quasi--stationary) state then
\begin{equation}
a(t)  \cong e^{\textstyle{ - \frac{i}{\hbar}({\cal E}_{\phi}^{0} -
\frac{i}{2} \gamma_{\phi}^{0})\,t }}. \label{a-q-stat}
\end{equation}
So, there is
\begin{equation}
{\cal E}_{\phi}^{0} - \frac{i}{2} \gamma_{\phi}^{0} \equiv i
\hbar\,\frac{\partial a(t)}{\partial t} \; \frac{1}{a(t)},
\label{E-iG}
\end{equation}
in the case of quasi--stationary states.

The standard interpretation and understanding of the quantum theory
and the related construction of our measuring devices are such that
detecting the energy ${\cal E}_{\phi}^{0}$ and decay rate
$\gamma_{\phi}^{0}$ one is sure that the amplitude $a(t)$ has the
form (\ref{a-q-stat}) and thus that the relation (\ref{E-iG})
occurs. Taking the above into account one can define the "effective
Hamiltonian", $h_{\phi}$, for the one--dimensional subspace of
states ${\cal H}_{||}$ spanned by the normalized vector
$|\phi\rangle$ as follows (see, eg. \cite{PRA})
\begin{equation}
h_{\phi} \stackrel{\rm def}{=}  i \hbar\, \frac{\partial
a(t)}{\partial t} \; \frac{1}{a(t)}. \label{h}
\end{equation}
In general, $h_{\phi}$ can depend on time $t$, $h_{\phi}\equiv
h_{\phi}(t)$. One meets this effective Hamiltonian when one starts
with the Schr\"{o}dinger Equation (\ref{Schrod}) for the total state
space ${\cal H}$ and looks for the rigorous evolution equation for
the distinguished subspace of states ${\cal H}_{||} \subset {\cal
H}$. In the case of one--dimensional ${\cal H}_{||}$  this rigorous
Schr\"{o}dinger--like evolution equation has the following form for
the initial condition $a(0) = 1$ (see \cite{PRA} and references one
finds therein),
\begin{equation}
i \hbar\, \frac{\partial a(t)}{\partial t} \;=\; h_{\phi}(t)\;a(t).
\label{eq-for-h}
\end{equation}
Relations (\ref{h}) and (\ref{eq-for-h}) establish a direct
connection between the amplitude $a(t)$ for the state $|\phi
\rangle$ and the exact effective Hamiltonian $h_{\phi}(t)$ governing
the time evolution in the one--dimensional subspace ${\cal H}_{\|}
\ni |\phi\rangle$. Thus the use of the evolution equation
(\ref{eq-for-h}) or the relation (\ref{h}) is one of the most
effective tools for the accurate analysis of the early-- as well as
the long--time properties of the energy and decay rate of a given
qausistationary state $|\phi (t) \rangle$.

So let us assume that we know the amplitude $a(t)$. Then starting
with this $a(t)$ and using the expression (\ref{h}) one can
calculate the effective Hamiltonian $h_{\phi}(t)$ in a general case
for every $t$. Thus, one finds the following expressions for the
energy and the decay rate of the system in the state $|\phi\rangle$
under considerations,
\begin{eqnarray}
{\cal E}_{\phi}&\equiv& {\cal E}_{\phi}(t) = \Re\,(h_{\phi}(t),
\label{E(t)}\\
\gamma_{\phi} &\equiv& \gamma_{\phi}(t) = -\,2\,\Im\,(h_{\phi}(t),
\label{G(t)}
\end{eqnarray}
where $\Re\,(z)$ and $\Im\,(z)$ denote the real and imaginary parts
of $z$ respectively.

As it was mentioned above the deviations of the decay law ${\cal
P}_{\phi}(t)$ from the exponential form can be described
equivalently using time-dependent decay rate. In terms of such
$\gamma_{\phi}(t)$ the Khalfin observation that ${\cal P}_{\phi}(t)$
must tend to zero as $t \rightarrow \infty$ more slowly than any
exponential function means that $\gamma_{\phi}(t) \ll
\gamma_{\phi}^{0}$ for $t \gg T_{\infty}$ and $\lim_{t \rightarrow
\infty} \,\gamma_{\phi}(t) = 0$.

The aim of this note is to examine the long time behaviour of ${\cal
E}_{\phi}(t)$ and $\gamma_{\phi}(t)$ using $a(t)$ calculated for the
given density $\omega({\cal E})$. We show that ${\cal E}_{\phi}(t)
\rightarrow 0$ as $t\rightarrow \infty$ for the model considered and
that a wide class of models has similar long time properties:
${{{\cal E}_{\phi}(t)}\vline}_{\;t \rightarrow \infty} \neq {\cal
E}_{\phi}^{0}$. It seems that in contrast to the standard Khalfin
effect \cite{Khalfin} in the case of the quasistationary states
belonging to the same class as excited atomic levels, this long time
properties of the energy ${\cal E}_{\phi}(t)$ have a chance to be
detected by analyzing spectra of very distant stars.

The paper is organized as follows. In Sec. 2 the above mentioned
model is considered. Sec. 3 contains an analysis of some  general
case of $a(t)$ causing ${\cal E}_{\phi}(t)$ to tend to zero as
$t\rightarrow \infty$. A discussion and final remarks can be found
in Sec. 4.

\section{The model}

Let us assume that ${Spec. (H)} = [0, \infty)$ and let us choose
$\omega ({\cal E})$ as follows
\begin{equation}
\omega ({\cal E}) = \frac{N}{2\pi}\,  \it\Theta ({\cal E}) \
\frac{\gamma_{\phi}^{0}}{({\cal E}-{\cal E}_{\phi}^{0})^{2} +
(\frac{\gamma_{\phi}^{0}}{2})^{2}}, \label{omega-BW}
\end{equation}
where $N$ is a normalization constant and
\[
\it\Theta ({\cal E}) \ = \left\{
  \begin{array}{c}
   1 \;\;{\rm for}\;\; {\cal E} \geq 0, \\
   0 \;\; {\rm for}\;\; {\cal E} < 0,\\
  \end{array}
\right.
\]
For such $\omega ({\cal E})$ using (\ref{a-spec}) one has
\begin{equation}
a(t) = \frac{N}{2\pi}  \int_{0}^{\infty}
 \frac{{\gamma_{\phi}^{0}}}{({\cal E}-{\cal E}_{\phi}^{0})^{2}
+ (\frac{\gamma_{\phi}^{0}}{2})^{2}}\, e^{\textstyle{ -
\frac{i}{\hbar}{\cal E}t}}\,d{\cal E}, \label{a-BW}
\end{equation}
where
\begin{equation}
\frac{1}{N} = \frac{1}{2\pi} \int_{0}^{\infty}
 \frac{\gamma_{\phi}^{0}}{({\cal E}-{\cal E}_{\phi}^{0})^{2}
+ (\frac{\gamma_{\phi}^{0}}{2})^{2}}\, d{\cal E}. \label{N}
\end{equation}
Formula  (\ref{a-BW}) leads to the result
\begin{eqnarray}
a(t) &=& N\,e^{\textstyle{- \frac{i}{\hbar} ({\cal E}_{\phi}^{0} -
i\frac{\gamma_{\phi}^{0}}{2})t}} \times \nonumber\\
&&\times \Big\{1 - \frac{i}{2\pi} \Big[
e^{\textstyle{\frac{\gamma_{\phi}^{0}t}{\hbar}}}\,
E_{1}\Big(-\frac{i}{\hbar}({\cal E}_{\phi}^{0}
+ \frac{i}{2} \gamma_{\phi}^{0})t\Big) \nonumber\\
&&\;\;\;\;\;+ (-1) E_{1}\Big(- \frac{i}{\hbar}({\cal E}_{\phi}^{0} -
\frac{i}{2} \gamma_{\phi}^{0})t\Big)\,\Big]\, \Big\}, \label{a-E(1)}
\end{eqnarray}
where $E_{1}(x)$ denotes the integral--exponential function
\cite{Sluis,Abramowitz}.

Using  (\ref{a-BW}) or (\ref{a-E(1)}) one easily finds that
\begin{equation}
i \hbar \,\frac{\partial a(t)}{\partial t} = ({\cal E}_{\phi}^{0} -
\frac{i}{2} \gamma_{\phi}^{0})\,a(t)\, + \,\Delta a(t), \label{Delta
-a}
\end{equation}
where
\begin{eqnarray}
\Delta a(t) &=& \frac{N}{\pi}\, \frac{\gamma_{\phi}^{0}}{2}
\,e^{\textstyle{- \frac{i}{\hbar}({\cal E}_{\phi}^{0} +
\frac{i}{2}\gamma_{\phi}^{0})t}}\times \nonumber \\
&& \;\;\;\;\;\;\;\;\times E_{1}\Big(-\frac{i}{\hbar}({\cal
E}_{\phi}^{0} + \frac{i}{2}\, \gamma_{\phi}^{0})t \Big).
\label{Delta-a-E1}
\end{eqnarray}
So,
\begin{equation}
h_{\phi}(t) \equiv i \hbar \,\frac{\partial a(t)}{\partial
t}\,\frac{1}{a(t)} \stackrel{\rm def}{=} h_{\phi}^{0} + \Delta
h_{\phi}(t), \label{h+Delta-h}
\end{equation}
where
\begin{equation}
h_{\phi}^{0} \equiv {\cal E}_{\phi}^{0} - \frac{i}{2}\,
\gamma_{\phi}^{0}, \label{h-0}
\end{equation}
and
\begin{equation}
 \Delta h_{\phi}(t)  =  +
\,\frac{\Delta a(t)}{a(t)}. \label{Delta-h}
\end{equation}

Using the asymptotic expansion of $E_{1}(x)$ \cite{Abramowitz},
\begin{equation}
{E_{1}(z)\vline}_{\, |z| \rightarrow \infty} \;\;\sim \;\;
\frac{e^{\textstyle{ -z}}}{z}\,( 1 - \frac{1}{z} + \frac{2}{z^{2}} -
\ldots ),  \label{E1-as}
\end{equation}
where $| \arg z  | < \frac{3}{2} \pi$, one finds
\begin{eqnarray}
{a(t)\vline}_{\, t \rightarrow \infty} &\simeq & N e^{\textstyle -
\frac{i}{\hbar}\,h_{\phi}^{0}\,t} \nonumber \\&& + (- i) \frac{N}{2
\pi}\; \frac{ \gamma_{\phi}^{0}}{|\,h_{\phi}^{0}\,|^{\,2}}
 \, \frac{\hbar}{t} \nonumber \\
&&+ (- 2)\,\frac{N}{2 \pi}\;\frac{{\cal E}_{\phi}^{0}\,
\gamma_{\phi}^{0}}{|\,h_{\phi}^{0}\,|^{\,4}} \,
\Big(\frac{\hbar}{t}\Big)^{2}\,+ \ldots\;\;, \label{a(t)-as}
\end{eqnarray}
and
\begin{eqnarray}
{\Delta a(t)\vline}_{\, t \rightarrow \infty} &\simeq & i\, \frac{N
\gamma_{\phi}^{0}}{2 \pi}\;
\frac{h_{\phi}^{0}}{|\,h_{\phi}^{0}\,|^{\,2}} \; \frac{\hbar}{t} \nonumber \\
&&+ \frac{N \gamma_{\phi}^{0}}{2 \pi}\;
\frac{(h_{\phi}^{0})^{2}}{|\,h_{\phi}^{0}\,|^{\,4}} \;
\Big(\frac{\hbar}{t}\Big)^{2}\;+\ldots\;\;. \label{Da(t)-as}
\end{eqnarray}
These two last asymptotic expansions  enable one to find (see
(\ref{Delta-h}))
\begin{eqnarray}
{\Delta h_{\phi}(t)\vline_{\,t \rightarrow \infty}} & = &
{\frac{\Delta a(t)}{a(t)} \,\vline}_{\;t \rightarrow \infty}
\nonumber \\ &\simeq& - \, h_{\phi}^{0} -\,i\,\frac{\hbar}{t}\,  -
\,2\, \frac{ {\cal E}_{\phi}^{0}}{ |\,h_{\phi}^{0} \,|^{\,2} } \Big(
\frac{\hbar}{t} \Big)^{2} +\ldots\;\;. \label{Delta-h-as}
\end{eqnarray}
Thus, there is
\begin{equation}
{h_{\phi}(t)\vline}_{\,t \rightarrow \infty} \simeq
-\,i\,\frac{\hbar}{t}\;  - \;2\, \frac{ {\cal E}_{\phi}^{0}}{
|\,h_{\phi}^{0} \,|^{\,2} }  \; \Big( \frac{\hbar}{t} \Big)^{2}
\;+\ldots \;\; \label{h-as}
\end{equation}
for the considered case (\ref{omega-BW}) of $\omega ({\cal E})$.

From (\ref{h-as}) it follows that
\begin{equation}
\Re\,({h_{\phi}(t)\vline}_{\,t \rightarrow \infty}) \;\stackrel{\rm
def}{=}\; {\cal E}_{\phi}^{\infty}\;
 \simeq \; -\,2\,
\frac{ {\cal E}_{\phi}^{0}}{ |\,h_{\phi}^{0} \,|^{\,2} }  \; \Big(
\frac{\hbar}{t} \Big)^{2} \, \longrightarrow\, 0,\label{Re-h-as}
\end{equation}
as $t \rightarrow \infty$, where ${\cal E}_{\phi}^{\infty} = {\cal
E}_{\phi}(t)|_{\,t \rightarrow \infty}$, and
\begin{equation}
\Im\,({h_{\phi}(t)\vline}_{\,t \rightarrow \infty}) \simeq
-\,\frac{\hbar}{t} \,\longrightarrow\, 0 \;\;\;({\rm as}\;\;\; t
\rightarrow \infty). \label{Im-h-as}
\end{equation}

For different states $|\phi_{1}\rangle, \,|\phi_{2}\rangle$ one has
\begin{eqnarray}
{\cal E}_{\phi_{1}}^{\infty} - {\cal E}_{\phi_{2}}^{\infty} &=&
-\,2\,\Big[ \frac{ {\cal E}_{\phi_{1}}^{0}}{ |\,h_{\phi_{1}}^{0}
\,|^{\,2} } \; -\; \frac{ {\cal E}_{\phi_{2}}^{0}}{
|\,h_{\phi_{2}}^{0} \,|^{\,2} } \;\Big]\; \Big( \frac{\hbar}{t}
\Big)^{2}\nonumber \\
&\neq& {\cal E}_{\phi_{1}}^{0} - {\cal E}_{\phi_{2}}^{0}\, \neq
0,\label{Re-h1-h2-as}
\end{eqnarray}
and
\begin{equation}
\Im\,({h_{\phi_{1}}(t)\vline}_{\,t \rightarrow \infty}) =
\Im\,({h_{\phi_{2}}(t)\vline}_{\,t \rightarrow
\infty}),\label{Im-h1-h2-as}
\end{equation}
whereas in general $\gamma_{\phi_{1}}\, \neq \,\gamma_{\phi_{1}}$.

The other relations are the following
\begin{eqnarray}
\frac{{\cal E}_{\phi_{1}}^{\infty}}{{\cal E}_{\phi_{2}}^{\infty}}
\,&=& \,\frac{{\cal E}_{\phi_{1}}^{0}}{{\cal E}_{\phi_{2}}^{0}}\;
\frac{ |\,h_{\phi_{2}}^{0} \,|^{\,2} }{
|\,h_{\phi_{1}}^{0} \,|^{\,2} } \nonumber\\
&=& \frac{{\cal E}_{\phi_{2}}^{0}}{{\cal E}_{\phi_{1}}^{0}}\; \frac{
1 + \frac{1}{4}\,(\frac{\gamma_{\phi_{2}}^{0}}{ {\cal
E}_{\phi_{2}}^{0}})^{2}}{1+
\frac{1}{4}\,(\frac{\gamma_{\phi_{1}}^{0}}{ {\cal
E}_{\phi_{1}}^{0}})^{2}} \;\;\neq\;\;\frac{{\cal
E}_{\phi_{1}}^{0}}{{\cal E}_{\phi_{2}}^{0}}, \label{Re-h1|Re-h2}
\end{eqnarray}
\begin{eqnarray}
\frac{{\cal E}_{\phi_{1}}^{\infty} - {\cal
E}_{\phi_{2}}^{\infty}}{{\cal E}_{\phi_{1}}^{\infty} + {\cal
E}_{\phi_{2}}^{\infty}} &=& \frac{\frac{{\cal
E}^{0}_{\phi_{1}}}{|h^{0}_{\phi_{1}}|^{2}} \,-\,\frac{{\cal
E}^{0}_{\phi_{2}}}{|h^{0}_{\phi_{2}}|^{2}}}{\frac{{\cal
E}^{0}_{\phi_{1}}}{|h^{0}_{\phi_{1}}|^{2}} \,+\,\frac{{\cal
E}^{0}_{\phi_{2}}}{|h^{0}_{\phi_{2}}|^{2}}} \nonumber\\
& \equiv & \frac{{\cal
E}^{0}_{\phi_{1}}\,|h^{0}_{\phi_{2}}|^{2}\,-\, {\cal
E}^{0}_{\phi_{2}}\,|h^{0}_{\phi_{1}}|^{2}}{{\cal
E}^{0}_{\phi_{1}}\,|h^{0}_{\phi_{2}}|^{2}\,+\, {\cal
E}^{0}_{\phi_{2}}\,|h^{0}_{\phi_{1}}|^{2}} \label{E1-E2|E1+E2}\\
& \neq & \frac{{\cal E}_{\phi_{1}}^{0} - {\cal
E}_{\phi_{2}}^{0}}{{\cal E}_{\phi_{1}}^{0} + {\cal
E}_{\phi_{2}}^{0}}, \nonumber
\end{eqnarray}
and
\begin{equation}
\frac{\Im\,({h_{\phi_{1}}(t)\vline}_{\,t \rightarrow
\infty})}{\Im\,({h_{\phi_{2}}(t)\vline}_{\,t \rightarrow
\infty})}\;=\;1\;\neq\;\frac{\gamma_{\phi_{1}}^{0}}{\gamma_{\phi_{2}}^{0}}.
\label{Im-h1|Im-h2}
\end{equation}
It seems  interesting that relations (\ref{Re-h1|Re-h2}) --
(\ref{Im-h1|Im-h2}) do not depend on time $t$.

Note that the following conclusion can be drawn from
(\ref{Re-h1-h2-as}):  For suitably long times $t$ there must be
\begin{equation}
\vline \; {\cal E}_{\phi_{1}}^{\infty}\, -\, {\cal
E}_{\phi_{2}}^{\infty}\,\vline \;\; <\;\; \vline\,{\cal
E}_{\phi_{1}}^{0} - {\cal E}_{\phi_{2}}^{0}\vline \,.
\label{E1-E2-as}
\end{equation}

These suitable times can be estimated using relation
(\ref{a(t)-as}). From (\ref{a(t)-as}) one obtains
\begin{equation}
{\vline\,{a(t)\vline}_{\, t \rightarrow \infty}\,\vline}^{\,2}
\simeq N^{2} e^{\textstyle - \frac{
\gamma_{\phi}^{0}}{\hbar}\,\,t}\; +\; \frac{N^{2}}{4
\pi^{2}}\;\frac{(\gamma_{\phi}^{0})^{2}}{|h_{\phi}^{0}\,|^{\,4}} \;
\frac{\hbar^{2}}{t^{2}}\; + \;\ldots\;\; . \label{t-as-1}
\end{equation}
Relations (\ref{Delta-h-as}) --- (\ref{E1-E2-as}) become important
for times $t > t_{as}$, where $t_{as}$ denotes the time $t$ at which
contributions to ${\vline\,{a(t)\vline}_{\, t \rightarrow
\infty}\,\vline}^{\,2}$ from the first exponential component in
(\ref{t-as-1}) and from the second component proportional to
$\frac{1}{t^{2}}$ are comparable. So $t_{as}$ can be be found by
considering the following relation
\begin{equation}
e^{\textstyle - \frac{ \gamma_{\phi}^{0}}{\hbar}\,\,t}\; \sim\;
\frac{1}{4
\pi^{2}}\;\frac{(\gamma_{\phi}^{0})^{2}}{|h_{\phi}^{0}\,|^{\,4}} \;
\frac{\hbar^{2}}{t^{2}}. \label{t-as-2}
\end{equation}
Assuming that the right hand side is equal to the left hand side in
the above relation one gets a transcendental equation.  Exact
solutions of such an equation can be expressed by means of the
Lambert $W$ function \cite{Corless}. An asymptotic solution of the
equation obtained from the relation (\ref{t-as-2}) is relatively
easy to find \cite{Olver}. The very approximate asymptotic solution,
$t_{as}$, of this equation for $(\frac{{\cal
E}_{\phi}}{\gamma_{\phi}^{0}})\,>\,10^{\,2}$ (in general for
$(\frac{{\cal E}_{\phi}}{\gamma_{\phi}^{0}})\,\rightarrow \,\infty$)
has the form
\begin{eqnarray}
\frac{\gamma_{\phi}^{0}\,t_{as}}{\hbar} &\simeq & 8,28 \,+\, 4\,
\ln\,(\frac{{\cal E}_{\phi}^{0}}{\gamma_{\phi}^{0}}) \nonumber \\
&&+\, 2\,\ln\,[8,28 \,+\,4\,\ln\,(\frac{{\cal
E}_{\phi}^{0}}{\gamma_{\phi}^{0}})\,]\,+\, \ldots \;\;.
\label{t-as-3}
\end{eqnarray}

\section{Some generalizations}

To complete the analysis performed in the previous Section let us
consider a more general case of $a(t)$. Namely, let the asymptotic
approximation to $a(t)$ have the form
\begin{equation}
a(t) \;\;
\begin{array}{c}
   {} \\
   \sim\\
   {\scriptstyle t \rightarrow \infty}
 \end{array}
\;\; \sum_{k=0}^{N} \,\frac{c_{k}}{t^{\lambda + k}},
\label{a(t)-as-w}
\end{equation}
where $\lambda > 0$ and $c_{k}$ are complex numbers. Note that the
asymptotic expansion for $a(t)$ of this or a similar form  one
obtains for a wide class of densities of energy distribution
$\omega({\cal E})$ \cite{Khalfin,Goldberger,Fonda,Peres,
Arbo,Dittes}.

From the relation (\ref{a(t)-as-w}) 
one concludes that
\begin{equation}
\frac{\partial a(t)}{\partial t} \;\;
\begin{array}{c}
   {} \\
   \sim\\
   {\scriptstyle t \rightarrow \infty}
 \end{array}
\;\; -\,\sum_{k=0}^{N} \,(\lambda + k)\,\frac{c_{k}}{t^{\lambda + k
+ 1}}. \label{da(t)-as-w}
\end{equation}

Now let us take into account the relation (\ref{eq-for-h}). From
this relation and relations (\ref{a(t)-as-w}), (\ref{da(t)-as-w}) it
follows that
\begin{equation}
h_{\phi}(t) \;\;
\begin{array}{c}
   {} \\
   \sim\\
   {\scriptstyle t \rightarrow \infty}
 \end{array}
\;\;\frac{d_{1}}{t}\, +
\,\frac{d_{2}}{t^{2}}\,+\,\frac{d_{3}}{t^{3}}\,+\,\ldots\,\,\, ,
\label{h-sim-w}
\end{equation}
where $d_{1}, d_{2}, d_{3}, \ldots$ are complex numbers. This means
that in the case of the asymptotic approximation to $a(t)$ of the
form (\ref{a(t)-as-w}) the following  property holds,
\begin{equation}
\lim_{t \rightarrow \infty}\,h_{\phi}(t) \, = \, 0. \label{lim-h}
\end{equation}

It seems to be important that  results (\ref{h-sim-w}) and
(\ref{lim-h}) coincide with the results (\ref{h-as}) ---
(\ref{Re-h1-h2-as}) obtained for the density $\omega ({\cal E})$
given by the formula (\ref{omega-BW}). This means that general
conclusion obtained for the other $\omega ({\cal E})$ defining
unstable states should be similar to those following from
(\ref{h-as}) --- (\ref{Re-h1-h2-as}).

\section{Final remarks.}

 Recently in many papers it was reported that some results of
 a detailed analysis of the spectra of very distant astrophysical object
could be explained by the variation of time of the fine structure
constant $\alpha$ \cite{Webb,Uzan,Bahcall,Reinhold}.
The analysis of the methods allowing  for such an interpretation of
the observed spectra shows that the conclusion that
$\alpha$ varies in time was drawn by comparing the emission lines of
two states separated by the fine-structure interactions
and obtained by observing an astrophysical object with the corresponding
lines coming form the laboratory source \cite{Bahcall}.
In the light of the results obtained in Sec. 2 and 3 the question arises
if such an interpretation of these astronomical observations is correct:
For the observer the effect described by relation (\ref{E1-E2|E1+E2})
and  caused by long time properties of time evolution of unstable states,
and the effect caused by a hypothetical variation in time of $\alpha$,
seems to be indistinguishable. Simply, there is a possibility that
the interpretation of the mentioned results of astronomical results
may not have taken into account all the available information. In the
following we will try to shed some light on the above statement.

So, let us consider a simple toy model. If one identifies energies
${\cal E}_{\phi}^{0}$ with energy emitted by an electron jumping
from the energy level $n_{j}$ to  the energy level $n_{k}$ of the
hydrogen atom in the nonrelativistic quantum mechanic then one finds
\begin{equation}
{\cal E}_{n_{jk}}^{0} = - R_{y} \,(\frac{1}{n_{j}^{2}} -
\frac{1}{n_{k}^{2}}), \;\;\;(n_{k}
 <  n_{j}),\label{R}
\end{equation}
where $n_{j}, n_{k}= 1,2,3, \ldots$ and $R_{y} = \frac{m_{e}
e^{4}}{2 \hbar^{2}} \equiv \frac{\alpha^{2} m_{e} c^{2}}{2}$ is the
Rydberg constant, $m_{e}$ is the mass of the electron, $c$ is the
speed of
light and $\alpha$ denotes the fine--structure constant.

Now let these atoms interact with some external fields and let these
fields excite the atoms forcing  electrons to change their energy
levels. Next, let these excited hydrogen atoms emit the
electromagnetic waves of the energies ${\cal E}_{n_{jk}}^{0} \equiv
h\,\nu_{n_{jk}}^{0}$ (where $\nu_{n_{jk}}^{0}$ denotes the frequency
of the emitted wave) and let them move away from the observer with
the velocity $v$. Then, the observer's measuring devices will show
that the energies emitted by these moving atoms are ${\cal
E}_{n_{jk}}^{0,\,v}$. The energy ${\cal E}_{n_{jk}}^{0}$ emitted by
this same source at rest is connected with ${\cal
E}_{n_{jk}}^{0,\,v}$ by the Doppler formula \cite{Heitler}
\begin{equation}
{\cal E}_{n_{jk}}^{0,\,v} = \kappa\;{\cal E}_{n_{jk}}^{0}\; <
\;{\cal E}_{n_{jk}}^{0}, \label{E-v}
\end{equation}
where $ \kappa =  \frac{1 - \beta}{\sqrt{1 - \beta^{2}}}$ and $\beta
= \frac{v}{c}$, and
\begin{equation}
{\cal E}_{n_{1k}}^{0,\,v} \mp {\cal E}_{n_{2k}}^{0,\,v} = \kappa\,
({\cal E}_{n_{1k}}^{0} \mp {\cal E}_{n_{2k}}^{0}). \label{E1-E2-v}
\end{equation}
>From (\ref{E-v}) and (\ref{E1-E2-v}) it follows that
\begin{equation}
\frac{{\cal E}_{n_{1k}}^{0,\,v}}{{\cal E}_{n_{2k}}^{0,\,v}} \equiv
\frac{{\cal E}_{n_{1k}}^{0}}{{\cal E}_{n_{2k}}^{0}},\;\;\;\; {\rm
or}\;\;\;\; \frac{\nu_{n_{1k}}^{0,\,v}}{\nu_{n_{2k}}^{0,\,v}} =
\frac{\nu_{n_{1k}}^{0}}{\nu_{n_{2k}}^{0}}, \label{E1|E2}
\end{equation}
where ${\cal E}_{n_{jk}}^{0,\,v}= h\,\nu_{n_{jk}}^{0,\,v}$, and
similarly
\begin{equation}
\frac{{\cal E}_{n_{1k}}^{0,\,v} - {\cal E}_{n_{2k}}^{0,\,v}}{{\cal
E}_{n_{1k}}^{0,\,v} + {\cal E}_{n_{2k}}^{0,\,v}} \,=\, \frac{{\cal
E}_{n_{1k}}^{0} - {\cal E}_{n_{2k}}^{0}}{{\cal E}_{n_{1k}}^{0} +
{\cal E}_{n_{2k}}^{0}}. \label{E1-E2|E1+E2-Dop}
\end{equation}

Now, let us assume for a moment that the $\omega({\cal E})$ given by
the formula (\ref{omega-BW}) describes the considered sources with
the sufficient accuracy.  Next if one additionally assumes that in
the distant past photons of energies, say ${\cal E}_{n_{jk}}^{0}$,
($j=1,2$), were emitted and that the sources of this emission were
moving away from the observer with known constant velocity $v$, then
at the present epoch this observer detects energies ${{\cal
E}_{n_{jk}}^{\infty\;}}' = \kappa\, {\cal E}_{n_{jk}}^{\infty}$,
where ${\cal E}_{n_{jk}}^{\infty} \stackrel{\rm def}{=}
\Re\,({h_{n_{jk}}(t)\vline}_{\,t \rightarrow \infty}$. If the moment
of this emission was at suitably  distant past then according to the
results of Sec 2 (see (\ref{E1-E2-as})) it should appear that
\begin{equation}
|{{\cal E}_{n_{1k}}^{\infty\;}}'\, -\, {{\cal
E}_{n_{2k}}^{\infty\;}}'\,|\;<\;|\, {{\cal E}_{n_{1k}}^{0,\,v}} -
{{\cal E}_{n_{2k}}^{0,\,v}}\,|\,. \label{E1-E2-infty}
\end{equation}

If this observer  knows nothing about the effect described in Sec. 2
and 3 and if he also finds that in addition
\begin{equation}
\frac{{\cal E}_{n_{1k}}^{\infty}}{{\cal E}_{n_{2k}}^{\infty}} \;
\neq\; \frac{{\cal E}_{n_{1k}}^{0}}{{\cal E}_{n_{2k}}^{0}},
\end{equation}
or
\begin{equation}
\frac{{\cal E}_{n_{1k}}^{\infty} - {\cal E}_{n_{2k}}^{\infty}}{{\cal
E}_{n_{1k}}^{\infty} + {\cal E}_{n_{2k}}^{\infty}} \,\neq\,
\frac{{\cal E}_{n_{1k}}^{0} - {\cal E}_{n_{2k}}^{0}}{{\cal
E}_{n_{1k}}^{0} + {\cal E}_{n_{2k}}^{0}}. \label{E1-E2|E1+E2-infty}
\end{equation}
then identifying ${{\cal E}_{n_{jk}}^{\infty\;}}'$ with ${{\cal
E}_{n_{jk}}^{0,\,v}}$, ($j=1,2$), and making reasonable assumptions
that $\frac{1}{\kappa}\,(\,{{\cal E}_{n_{1k}}^{\infty\;}}'\, -\,
{{\cal E}_{n_{2k}}^{\infty\;}}'\,)$ should also fulfil the relation
of the type (\ref{R}), and the ratio $\frac{{{\cal
E}_{n_{1k}}^{\infty\,}}'}{{{\cal E}_{n_{2k}}^{\infty\,}}'} \,\equiv
\,\frac{{\cal E}_{n_{1k}}^{\infty}}{{\cal E}_{n_{2k}}^{\infty}}$ or
$\frac{{\cal E}_{n_{1k}}^{\infty} - {\cal
E}_{n_{2k}}^{\infty}}{{\cal E}_{n_{1k}}^{\infty} + {\cal
E}_{n_{2k}}^{\infty}}$ should be the same as that at the present
epoch (see (\ref{E1|E2})\,), this observer, on the basis of property
(\ref{E1-E2-as}), is forced to conclude that in the mentioned
suitably distant past the Rydberg constant $R_{y}^{past}
\stackrel{\rm def}{=} R_{y}|_{\,t\rightarrow \infty}$ was smaller
than at the present epoch.

On the other hand, if the observer knows the effect described in the
previous Sections then he could  attribute the decrease of $R_{y}$
in the distant past to the long time behavior of the survival
probability $a(t)$ and resulted from this the decrease of the
effective Hamiltonian $h_{\phi}(t)$ for $ t \rightarrow \infty$ (see
(\ref{h}), (\ref{h-as}), (\ref{Re-h-as}) and (\ref{h-sim-w})\,)
rather than to an actual change in the fine structure constant. The
conclusion that such an observation can be considered as a
confirmation of the mentioned effect seems to be justified. Note
that from (\ref{E1|E2}), (\ref{E1-E2|E1+E2}) it follows that this
effect seems to be distinguishable from the Doppler effect.

Unfortunately, for the reasons mentioned in Sec. 1 it seems to be
rather
impossible to observe the discussed effect in laboratory tests. A
similar conclusion follows from the relations (\ref{t-as-2}),
(\ref{t-as-3}). Nevertheless, this effect should be detectable by
means of the observation of spectra of distant astrophysical object.

As mentioned earlier
some astrophysical observations suggest that relations of type
(\ref{E1-E2-infty}) and (\ref{E1-E2|E1+E2-infty})  take place. More
precisely, these observations suggest that at very distant past the
fine structure constant $\alpha$ was smaller than at the present
epoch (for the recent data see \cite{Webb}) which implies that
$R_{y}^{past}\,<\,R_{y}$. The known physical effects responsible for
the observed redshift are unable to explain such a result
\cite{Kolb}. Therefore, the hypothesis that $\alpha$ varies in time,
$\alpha = \alpha (t)$, and that at the very distant past $\alpha
(t)$ was smaller than $\alpha (0)$ (where $\alpha (0)$ denotes the
value of $\alpha (t)$ at the present epoch) was formulated (see
\cite{Webb,Uzan,Bahcall} and references one can find therein). The
scale of changes of $\alpha$ is usually described by means of the
quantity $\frac{\Delta \alpha (t)}{\alpha (0)}$, where: $\Delta
\alpha (t) = \alpha (t) - \alpha (0)$.

One of the methods allowing to estimate $\frac{\Delta \alpha
(t)}{\alpha (0)}$ consists in the analysis of the emission lines of
two states separated by the fine--structure interactions. In such a
case to a very good accuracy (see formula (5) in \cite{Bahcall})
\begin{equation}
\Big( \frac{\alpha (t)}{\alpha (0)}\Big)^{2} \;\simeq\;
\Big[\;\frac{\lambda_{2}(t) - \lambda_{1}(t)}{\lambda_{2}(t) +
\lambda_{1}(t)}\Big] \; \Big[\;\frac{\lambda_{2}(0) +
\lambda_{1}(0)}{\lambda_{2}(0) - \lambda_{1}(0)}\Big], \label{a2|a2}
\end{equation}
where $\lambda_{j}(t)$, ($j = 1,2$), are wavelengths at time $t$.
Using the relation ${\cal E}_{n_{jk}}(t) =
\frac{hc}{\lambda_{j}(t)}$, ($j = 1,2$),  one can rewrite this
relation as follows
\begin{eqnarray}
\Big( \frac{\Delta \alpha (t)}{\alpha (0)}\;+\;1 \Big)^{2} &\simeq &
\Big[\;\frac{{\cal E}_{n_{1k}}(t) - {\cal E}_{n_{2k}}(t)}{{\cal
E}_{n_{1k}}(t) + {\cal E}_{n_{2k}}(t)}\Big] \times \nonumber\\
&&\times \Big[\;\frac{{\cal E}_{n_{1k}}(0) + {\cal
E}_{n_{2k}}(0)}{{\cal E}_{n_{1k}}(0) - {\cal E}_{n_{2k}}(0)}\Big].
\label{da|a0}
\end{eqnarray}
Taking $t \rightarrow \infty$ in this formula and replacing ${\cal
E}_{n_{jk}}(t)$, ($j = 1,2$), by relations of type (\ref{E(t)}),
(\ref{Re-h-as}), i. e., by ${\cal E}_{n_{jk}}^\infty$, one obtains
an expression for ${\frac{\Delta \alpha (t)}{\alpha
(0)}\vline}_{\;t\rightarrow \infty}$ connecting the variation of
$\alpha$ with the effect described in Sec. 2.  Of course the
contribution of this effect into the possibly non-zero value of
$\frac{\Delta \alpha (t)}{\alpha (0)}$ is only connected with the
long time behavior of the survival amplitude $a(t)$, that is  with
the long time properties of the  real part of the effective
Hamiltonian, ${\cal E}_{n_{jk}}(t) = \Re\,(h_{n_{jk}}(t))$,
described earlier and does not signify the change of $\alpha$ with
time.

Note that from (\ref{E1-E2|E1+E2}) and other results of Sec. 2 it
follows that the right hand side of (\ref{da|a0}) should not be
equal to 1 for $t \rightarrow \infty$. So if one tries to estimate
possible variations of $\alpha$ in time $t$ using data obtained from
astrophysical observation of a very distant cosmic object, one
should take into account that a significant contribution to
$\Delta\alpha$ obtained in this way can be a contribution generated
by the long-time effect discussed in Sec 2.

Unfortunately, the model considered in Sec. 2 does not reflect
correctly all properties of the real physical system containing
unstable states. Therefore, formula (\ref{E1-E2|E1+E2}) obtained
within this model can not be considered as universally valid and
it does not lead to the expression for $\frac{\Delta \alpha
(t)}{\alpha (0)}$ correctly describing real properties of very
distant astronomical sources of electromagnetic radiation. The
defect of this model is that some quantities calculated within it
are divergent. Indeed from (\ref{a(t)}), (\ref{a-spec}) and
(\ref{Schrod}) one finds that
\begin{equation}
i \hbar\,{{\frac{\partial a(t)}{\partial t}}\vline}_{\,t=0} =
\langle\phi|H|\phi\rangle, \label{da|dt-t=0}
\end{equation}
but inserting into (\ref{a-spec}) the density $\omega ({\cal E})$
given by (\ref{omega-BW}),  relation (\ref{da|dt-t=0}) yields
$\langle\phi|H|\phi\rangle = \infty$. For this same reason
 $h_{\phi}(t)$ and $\Delta h_{\phi}(t)$ computed for
$\omega ({\cal E})$ given by (\ref{omega-BW}) are divergent at
$t=0$: $ h_{\phi}(=0) = \infty, \;\; \Delta h_{\phi}(t=0) = \infty
$. Nevertheless, taking into account that a very wide class of
models of unstable states leads to a similar asymptotic long time
behaviour of $a(t)$ to this obtained for the model considered in
Sec. 2, some general conclusions following from the results of Sec.
2 and 3 seem to deserve more attention.

So, the following conclusion which can be drawn from
(\ref{E1-E2|E1+E2}) seems to hold: If one considers the same pairs
of spectral lines (see formula (92) in \cite{Uzan}, or (5) in
\cite{Bahcall}) then the above discussed contribution into $\Delta
\alpha$ calculated with  the use of astrophysical data obtained for
suitably distant two different sources, such that the second source
is much older (i.e. much more distant) than the first one, can be
the same for these different sources. What is more (see
(\ref{E1-E2|E1+E2})), this contribution need not depend on time $t$
but it should depend on such parameters describing the excited
energy levels of atoms emitting the registered electromagnetic
radiation and corresponding to measured spectral lines as energies
${\cal E}^{0}_{n_{jk}}$ and decay rate $\gamma^{0}_{n_{jk}}$ (or
lifetimes). This means that estimations of $\Delta \alpha$ performed
by means of relations of type (\ref{a2|a2} \textit{- OK}), (\ref{da|a0}) and
using the astrophysical data obtained for the given cosmic source
emitting the electromagnetic radiation can lead to different values
of $\Delta \alpha$ depending on the pairs of spectral lines used for
calculations. If the hypothesis that $\alpha$ varies in time is true
then it seems to be obvious that the above discussed contribution
into $\Delta \alpha$ should be absent when one tries to estimate
possible variations of $\alpha$ in time in laboratory test.
Therefore one should expect the values of $\Delta \alpha$ obtained
in laboratory tests \cite{Uzan} can differ from those obtained using
astrophysical data. Simply the magnitude of possible variations of
$\alpha$ obtained within the use of astrophysical data can be
enhanced by the contribution of the effect discussed in this paper.

All the above remarks concerning estimations  of $\Delta \alpha$
using results of observations of the spectra of distant astrophysical
objects hold also when one tries to indicate a possible cosmological
variation of the proton--electron mass ratio analyzing  spectra of
the same objects \cite{Uzan,Reinhold}.
Cosmic distances and  other parameters computed
from the observed redshift of very distant objects emitting
electromagnetic radiation \cite{Kolb} are calculated without taking
into account the possible quantum long time energy redshift
described in Sec. 2 and Sec. 3, so these distances as well as values
of these parameters need not
reflect real values of these parameters and thus need not reflect
correctly real properties of the Universe.

At this point it should probably be made clear that the discussion
of this Section and the results obtained in Sec. 2 and 3 are not
meant to provide an alternative explanation of the source of the
potential variation of $\alpha$. Rather we point out there is
a possibility to go beyond the standard interpretation of
some results of the measurements of the scale of this potential variation.


\begin{thebibliography}{10}
\bibitem{Fock}
S. Krylov, V. A. Fock, Zh. Eksp. Teor. Fiz. {\bf 17}, (1947), 93.
\bibitem{Khalfin}
L. A. Khalfin, Zh. Eksp. Teor. Fiz. {\bf 33}, (1957), 1371 [Sov.
Phys. --- JETP {\bf 6}, (1958), 1053].
\bibitem{Paley}
R. E. A. C. Paley, N. Wiener, {\em Fourier transforms in the comlex
domain}, American Mathematical Society, New York, 1934.
\bibitem{Goldberger}
M. L. Goldberger, K. M. Watson, {\em Collision Theory}, Willey, New
York 1964.
\bibitem{Newton}
R. G. Newton, {\em Scattering Theory of Waves and Particles},
Springer, New York 1982.
\bibitem{Fonda}
L. Fonda, G. C. Ghirardii and A. Rimini, Rep. on Prog. in Phys. {\bf
41}, (1978), 587.
\bibitem{Peres}
A. Peres, Ann. Phys. {\bf 129}, (1980), 33.
\bibitem{Greenland}
P. T. Greenland, Nature {\bf 335}, (1988), 298.
\bibitem{Arbo}
D. G. Arbo, M. A. Castagnino, F. H. Gaioli and S. Iguri, Physica
{\bf A 227}, (2000), 469.
\bibitem{PRA}
K. Urbanowski, Phys. Rev. {\bf A 50}, (1994), 2847.
\bibitem{Wessner}
J. M. Wessner, D. K. Andreson and R. T. Robiscoe, Phys. Rev. Lett.
{\bf 29}, (1972), 1126.
\bibitem{Norman1}
E. B. Norman, S. B. Gazes, S. C. Crane and D. A. Bennet, Phys. Rev.
Lett. {bf 60}, (1988), 2246.
\bibitem{Sluis}
K. M. Sluis, E. A. Gislason, Phys. Rev. {\bf A 43}, (1991), 4581.
\bibitem{Abramowitz}
{\em Handbook of Mathematical Functions}, Natl. Bur. Stand. Appl.
Math. Ser. No 55, eds. M. Abramowitz nad I. A. Stegun (U.S. GPO,
Washington, D.C., 1964).
\bibitem{Corless}
R. M. Corless, G. H. Gonet, D. E. G. Hare, D. J. Jeffrey and D. E.
Khnut, Adv. Comput. Math. {\bf 5}, (1996), 329.
\bibitem{Olver}
F. W. J. Olver, {\em Asymptotics and special functions}, Academic
Press, New York, 1974. 
\bibitem{Dittes}
F. M. Dittes, H. L. Harney and A. M\"{u}ller, Phys. Rev. {\bf A 45},
(1992), 701.
\bibitem{Heitler}
W. Heitler, {\em The Quantum Theory of Radiation}, Oxford University
Press, London 1954 and Dover Publications, New York 1984.
\bibitem{Webb}
J. K. Webb, M. T. Murphy, V. V. Flaubaum, V. A. Dzuba, J. D. Barrow,
C. W. Churchill, J. X. Prochaska and A. M. Wolfe, Phys. Rev. Lett.
{\bf 87}, (2001), 091301; J. Magueijo, J. D. Barrow, H. B. Sandvik,
Phys. Lett. {\bf B 549}, (2002), 284; M. T. Murphy, J. K. Webb, V.
V. Flaubaum, Mon. Not. R. Astron. Soc. {\bf 345}, (2003), 609; R.
Srianand, H. Chand, P. Petijean and B. Brasil, Phys. Rev. Lett. {\bf
92}, (2004), 121302; E. Peik, B. Lipphardt, H. Schneider and Chr.
Tamm, Phys. Rev. Lett. {\bf 93} (2004), 170801; C. J. A. P. Martins,
A. Melchiorri, G. Rocha, R. Trotta, P.P. Avelino and P. T. P. Viana,
Phys. Lett. {\bf B 585}, (2004), 29; P. Tzanavaris, J. K. Webb, M.
T. Murphy, V. V. Flaubaum and S. J. Curran, Phys. Rev. Lett. {\bf
95} (2005), 041301.
\bibitem{Kolb}
E. W. Kolb, M. S. Turner, {\em The Early Universe}, Addison--Wesley
Publ. Co., 1993.
\bibitem{Uzan}
J.--P. Uzan, Rev. Mod. Phys. {\bf 75}, (2003), 403.
\bibitem{Bahcall}
J. N. Bahcall, Astrophys. J. {\bf 600}, (2004), 520.
\bibitem{Reinhold}
E.Reinhold, R.Buning, U. Hollenstein, A. Ivanchik, P. Petijean and
W. Ubachs, Phys. Rev.  Lett. {\bf 96}, (2006), 151101.
\end{thebibliography}
\end{document}